\documentstyle[graphicx]{jaa}

\begin{document}

\title[Binary black hole search from VLBI images]{A search for binary black hole candidates from the VLBI images of AGNs}
\author[Xiang Liu]%
       {Xiang Liu$^{1,2,*}$\\
        $^1$ Xinjiang Astronomical Observatory, CAS, 150 Science 1-Street, Urumqi 830011, China\\
        $^2$ Key Laboratory of Radio Astronomy, CAS, Nanjing 210008, China\\
        $^*$ email: liux@xao.ac.cn\\}

\maketitle

\label{firstpage}

\begin{abstract}
We have searched the core-jets pairs in the VLBI scales ($<$1
kpc), from several VLBI catalogues, and find out 5 possible binary
black hole (BBH) candidates. We present the search result and
analyze the candidates preliminarily, and further study with
multi-band VLBI observation and optical line investigation would
be carried out in the future.
\end{abstract}

\begin{keywords}
Binary black hole -- radio continuum: jets pair
\end{keywords}

\begin{centering}
\section{Introduction}
\end{centering}
In the co-evolution scenario of galaxies and their supermassive
black holes, galaxy-galaxy mergers would end up with forming
binary black holes (BBHs). Close binary black holes are important
in astrophysics and also they would be the strongest gravitational
wave sources in the universe. As the binary orbiting black holes
give off gravitational waves, their orbit decays and the orbital
period decreases. This stage is called binary black hole inspiral.
Galaxy-galaxy merging systems are mostly found in optical and
X-ray images. In radio, it is less efficient in identifying active
nuclei pairs of merging systems, because most of them are
radio-quiet. However, an advantage in radio is the VLBI which can
resolve the close binary system at pc scale, if the nuclei are
radio loud. In $>$1 kpc scales, the frequency of double nuclei of
merging galaxies was estimated to be $\sim$1\% in the optical
samples (Wang, et al. 2009; Liu et al. 2013). Assuming 10\% of the
AGNs are radio loud in optical sample, the detection rate of
twin-jets pairs in radio reduces to $\sim$0.1\% in the optical
selected sample. A similar fraction ($\sim$0.1\%) is expected in
the radio selected sample for both the double nuclei being radio
loud.

The VLBI can image the radio jets of AGN at pc scales, providing
an unprecedented high resolution probe to close binary black
holes. Even if the detection rate as low as $\sim$0.1\%, with
up-to-date large VLBI databases one could still be able to find
out close binary radio jets pairs launched from binary black
holes. Such a VLBI search for close binary radio cores done from
the geodetic VLBI database, found only one binary cores system
from 3114 sources (Burke-Spolaor 2011). The detection rate seems
less than that estimated above by a factor of 3. We have searched
the core-jets pairs from astrophysical databases, and present the
preliminary results in this paper.\\

\begin{centering}
\section{Radio jets pairs as the BBH signatures}
\end{centering}
It is more efficient to find binary AGNs in $<$10 kpc than in 10
-100 kpc scales, suggesting that simultaneous fuelling of both
black holes is more common as the binary orbit decays through
shorter spacings (Smith et al. 2010). If both nuclei are radio
loud, in $<$1~kpc scales, the VLBI is the most efficient probe to
find the compact inspiral cores. However, in a relatively small
sample of AGNs with double-peaked optical emission lines, Tingay
\& Wayth (2011) have not detected compact double cores with the
VLBA. Burke-Spolaor (2011) searched for flat spectrum double cores
in large geodetic VLBI database, and find only one, i.e.
B0402+379. The apparent deficit of double cores at the small
spacings is not clear, a reason may be that one of double cores
could be obscured/absorbed or very weak. Independent searches with
different strategies from different samples are needed to confirm
whether the deficit is real.

Our strategy of searching for BBH candidates is looking for not
only flat spectrum double cores, but also twin-jets pair. Double
twin-jets are found more frequently in $>$1 kpc scales than in the
VLBI scale, e.g. the 19 X-shaped radio sources and 100 X-shaped
candidates (Cheung, 2007), and the double-double radio galaxies
(e.g. Liu et al. 2003), whereas their radio cores are often absent
or weak probably due to large viewing angles. The VLBI databases
in our searches are the MOJAVE database (Lister et al. 2009), the
Pearson-Readhead survey and the first Caltech-Jodrell Bank survey
(PR+CJ1, e.g. Xu et al. 1995), the second Caltech-Jodrell Bank
survey (CJ2, e.g. Henstock et al. 1995), the CJF sample (Britzen
et al. 2007), the VLBApls sample (Fomalont et al. 2000), and the
VIPS sample (Helmboldt et al. 2007).
The resulted candidates from $\sim$2000 sources are summarized in Table 1. We define a spectral index with $S \propto \nu^{\alpha}$.\\

\begin{table*}
         \caption[]{The 6 BBH candidates, the number in the last column are catalogues: 1--MOJAVE, 2--CJ1, 3--CJ2, 4--CJF, 5--VLBApls, 6--VIPS}
         $$
         \begin{tabular}{ccccc}
\hline
  \hline
    \noalign{\smallskip}
    Name & Other & Id & z      &  Cat  \\
\hline
  \noalign{\smallskip}

J0405+3803  & B2 0402+379  &  G  &  0.055046          & (4,5)     \\

J1001+5540  & 4C55.19      &  G  & 0.003723           & (6)       \\

J1158+2450  & PKS 1155+251 &  Q  &  0.201600          & (1,5,6)   \\

J1215+3448  & B2 1213+350  &  Q  & 0.857000           &  (2,4,5,6)  \\

J1632+3647  & B2 1630+35   &     &                    & (6)        \\

J2253+1608  & 3C454.3     &  Q   &  0.859000          &  (1,5)    \\

           \noalign{\smallskip}
            \hline
           \end{tabular}{}
         $$
         \label{tab1}
   \end{table*}

\begin{centering}
\section{Comments on individual BBH candidates}
\end{centering}
We find 6 possible BBH candidates from our intensive searches,
they have shown features likely either double core-jets or a
compact core plus a core-jet. We comment and analyze preliminarily
these sources in the following.

B2 0402+379

This is a previously suggested supermassive BBH system which
showing a compact radio core and a nearby Compact Symmetric Object
(Rodriguez et al. 2006). Two components of HI absorption lines
were found, which are supporting the BBH model of the source
(Morganti et al. 2009).

4C55.19 (J100157.93+554047.8)

The host is a nearly edge-on spiral galaxy NGC 3079 with a LINER
nucleus, which is classified as a Seyfert 2 galaxy and a low
luminosity AGN. The VIPS image at 5 GHz exhibits three compact
components, the southeast one (A) and the northwest one (B) are
consistent with the detections with global VLBI by Sawada-Satoh et
al. (2000, SS hereafter). The third component in the northeast of
the VIPS image has a position angle close to the minor axis of the
galaxy, which may either relate to a fresh outflow, or might be a
young supernova remanent, this component was not detected in SS.
Mostly likely the component B is the core and A is a jet as
suggested by SS, but the component A is more compact and brighter
than B in the 5 GHz VIPS image, indicating a strong variability of
the components compared with the 5 GHz image of Trotter et al.
(1998). The jet is misaligned with the short axis of the galaxy,
the radio `core' is at about 0.5 pc west to the galactic nucleus
according to Trotter et al. (1998), implying this radio core is
probably a secondary core in a BBH system. Hagiwara et al. (2004)
found two OH absorption components in the central region, but its
relation to a possible BBH system needs further investigation.

PKS 1155+251 (SDSS J11584+2450)

The quasar is point like in both the NVSS and FIRST images, and
shows a slightly extension along southeast-northwest with the VLA
D-array (Tremblay et al. 2008) and they reclassified this source
as a CSO, and discussed a possible shrinking of the source from
new VLBA observations at 4.84, 8.34, and 15.13~GHz. However, the
complex structure of the source was not fully explained with the
CSO scenario, for the exotic diffuse emission towards the west
which is perpendicular to the CSO jet axis. The 15 GHz images in
Tremblay et al. (2008) and in the MOJAVE showed the southern
brightest component is quite compact, although its spectral index
being steep as estimated from the 15 GHz and lower frequencies
(but note that the compact component embedded in the jet/diffuse
emission which may lead to an overestimate its flux densities at
the lower frequencies), a spectral index between 15 and 43 GHz is
vital to clarify its nature. The spectrum of the source total flux
density shows a flat spectrum in $<$3~GHz in the NED, this flat
spectrum is not consistent with a usual steep spectrum of CSO.
Furthermore, there seems a jet in the west connected to the
southern-end component. Therefore, possibly the southern-end
component is a core. In this scenario, the source has two cores
and jets. The diffuse emission in the west could be due to the
precession of the binary system. We measured the separation
between the two candidate cores with the MOJAVE data at three
epochs 1995.4, 1999.5 and 2001.3, the resulted separation is
within 3.38$\pm$0.13 mas, in supporting the non-expansion between
the central core component and the southern brightest component
(Tremblay et al. 2008). Kellermann et al. (2004) also considered
this complex source had two split jets. We keep the source as a
possible BBH candidate for future VLBI monitoring and optical line
study.

B2 1213+350 (SDSS J121555.60+344815.2)

The quasar shows a point like in the NVSS image, while has a short
jet to northeast in the FIRST image. The VLBI images at 1.6 and 5
GHz from the CJ1 showed a strong hotspot-jet association in the
northern and a compact component in the south. The 5 GHz VIPS
image revealed a short jet from the southern `core' candidate.
Likely the southern one is the core and the northern part is the
jet/hotspot. However, the northern feature seems being jetted to
southeast and then the jet curved to northeast. The overall radio
spectrum of the source is -0.35 (1.4 GHz to 30 GHz) from the NED.
From our estimate, the spectral indices of both the northern
bright component and the southern core are flat from the 2.3/8.4
GHz RRFID VLBI data and the 5 GHz VLBApls data, so an alternative
scenario could be still possible that the northern part harbors a
secondary core. We keep it as a BBH candidate for further
observations and study in detail, high frequency VLBI observation
at 43 GHz are very needed to classify this source.

B2 1630+35

No optical and high energy band information available for this
source, the radio spectrum shows a peaked shape around 1 GHz -- a
GPS source (Marecki et al. 1999). Both the NVSS and FIRST images
show unresolved. The 5 GHz VIPS image exhibits a northern compact
component and southeastern core-jet like features. No VLBI images
at other frequency are available for a spectral index estimate of
the components. Multi-band high resolution VLBI observations
should be able to classify the source.

3C454.3

This is a highly variable quasar through whole electromagnetic
spectrum. Total flux density has a slightly peaked spectrum around
1 GHz. The kpc scale radio image indicates a core-jet to
northwest. The 15 GHz MOJAVE images revealed overall core-jets
aligning with the kpc scale jet. The VLBA features are complex,
with the first part: the east-end core-jet with diffuse ring
emission widely around the jet within 5 mas, and the second part:
a possible secondary `core'-jet like feature beyond 5 mas in the
northwest. We measured the secondary `core' position with respect
to the primary core from the MOJAVE data, find that the angular
distance is stable with 6.0$\pm$0.5 mas in the past 17 years. The
source was suggested to be a BBH system by Britzen et al. (2012),
for its ring-like feature around the primary core-jet. We have a
different view that the secondary BH might be locating at 6 mas
northwest of the primary core, in this scenario, the ring-like
feature may be resulted from a kind of reflection of the primary
jets by the accretion disk of the secondary BH, more detailed
modelling will be giving in
the future.\\

\section*{\centering Acknowledgements}
The work is supported by the 973 Program of China (2009CB824800)
and the NSFC grant 11273050. This research has made use of data
from the MOJAVE database that is maintained by the MOJAVE team
(Lister et al. 2009). This research has made use of the United
States Naval Observatory (USNO) Radio Reference Frame Image
Database (RRFID).

\end{document}